\documentclass[prb,twocolumn,showpacs,floatfix]{revtex4}

\usepackage{graphics}
\usepackage{graphicx}
\def  \bsig   {\mbox{\boldmath$\sigma$}}

\def \bmatA   {\mbox{\boldmath$\mathcal{A}$}}

\def \bta      {\mbox{\boldmath$\tau$}}

\begin{document}

\title{Graphene in periodically alternating magnetic field: 
unusual quantization of the anomalous Hall effect}

\author{M. Taillefumier$^{1}$, V. K. Dugaev$^{2,3,4}$, B. Canals$^4$,
C. Lacroix$^4$, and P. Bruno$^5$}

\affiliation{$^1$Department of Physics, University of Oslo, 0316 Oslo, Norway\\
$^2$Department of Physics, Rzesz\'ow University of Technology,
al. Powsta\'nc\'ow Warszawy 6, 35-959 Rzesz\'ow, Poland\\
$^3$Department of Physics and CFIF, Instituto Superior T\'ecnico,
TU Lisbon, Av. Rovisco Pais, 1049-001 Lisbon, Portugal\\
$^4$Institut N\'eel, CNRS/UJF, 25 avenue des Martyrs, BP 166,
38042 Grenoble, Cedex 09, France\\ 
$^5$European Synchrotron Radiation Facility, BP 220, 38043 Grenoble, Cedex, France}

\date{\today }

\begin{abstract}
  We study the energy spectrum and electronic properties of graphene
  in a periodic magnetic field of zero average with a symmetry of
  triangular lattice.  The periodic field leads to formation of a set
  of minibands separated by gaps, which can be manipulated by
  external field. The Berry phase, related to the motion of electrons
  in $k$ space, and the corresponding Chern numbers characterizing
  topology of the energy bands are calculated analytically and
  numerically. In this connection, we discuss the anomalous Hall
  effect in the insulating state, when the Fermi level is located in
  the minigap. The results of calculations show that in the model of
  gapless Dirac spectrum of graphene the anomalous Hall effect can be
  treated as a sum of fractional quantum numbers, related to the
  nonequivalent Dirac points.
\end{abstract}
\maketitle

\section{Introduction}

The energy structure of graphene includes the Dirac points where the energy spectrum of electrons 
is relativistic. This can lead to many very unusual 
transport properties of this
material \cite{novoselov04,novoselov05,geim07,neto09,peres10}. One of
examples is the unusual integer quantum Hall effect in graphene.
As found theoretically and confirmed experimentally
\cite{novoselov05,zhang05} there is no $N=0$ plateau, which is
related to the Dirac point in the electronic spectrum. 
The other very important property is that the
Landau splitting is very strong in graphene, which makes it possible
to observe the quantum Hall effect at room temperatures
\cite{novoselov07}.

Recently, we proposed to use a periodic magnetic field for
quantization of the electron energy spectrum of two-dimensional
electron gas \cite{taillefumier08}. Our calculations demonstrated that
periodically alternating field leads to an energy spectrum with a
number of minibands separated by energy gaps. In such system,
the anomalous Hall effect (AHE) is nonzero, and it is quantized if the
chemical potential is located within the gap.
It was also suggested \cite{bruno04,taillefumier08} that the
realization of such structure under periodic magnetic field can be
easily achieved by using a lattice of magnetic nanorods
\cite{nielsch01}. It should be noted that another possible way is to 
use the 2D skyrmion lattice \cite{rossler06}, which have been recently
observed in thin layers of helical magnets
(however, in this case the temperature should be rather low) \cite{yu10}.

Graphene has obvious advantages to be used instead of semiconducting
quantum well with two-dimensional electron spectrum. Graphene is
naturally two-dimensional, and the technology of graphene-based
structures is much simpler. Besides, one can expect much stronger
effects related to the magnetic quantization due to the specific
parameters of graphene.
      
In this work we calculate the electron energy spectrum and the Chern
numbers characterizing topological properties of the energy bands. The
results are essentially different for graphene as compare to the 2D
electron gas with parabolic energy spectrum. We find that the Chern
numbers of the bands corresponding to higher-in-energy excitations of
electrons and holes are zero at sufficiently large field. On the
contrary, the low-energy bands of electrons and holes have nonzero
Chern numbers, which results in the quantized anomalous Hall effect.

\section{Generalized model with the gap}

We start with a generalized model, which includes the relativistic
Hamiltonian \cite{neto09} describing the electron energy spectrum near
the Dirac points $\mathcal{K,\, K'}$ in the Brillouin zone of graphene
(Fig.~1a) and an additional term leading to the gap
\begin{eqnarray}
\label{1}
\mathcal{H}^{\mathcal{K,\, K'}}
=
\mp iv\tau _x\left( \partial _x-\frac{ieA_x}{c}\right) 
-iv\tau _y\left( \partial _y-\frac{ieA_y}{c}\right)
\nonumber \\
+\Delta \tau _z,
\end{eqnarray}
where the components of vector $\bta $ are the Pauli matrices acting
in the space of graphene sublattices.
Introducing the gap parameter $\Delta $ makes the A and B
sites different in the crystal lattice of graphene.  Physically, it can be the
case of different atoms in the honeycomb lattice like in
two-dimensional boron nitride. The model with $\Delta \ne 0$ can be
also realized for graphene in periodic electric and magnetic fields
discussed in Ref.~[\onlinecite{snyman09}].  In the case of graphene we should
put $\Delta =0$ but for pedagogical reasons it is instructive to
start from $\Delta \ne 0$.
The vector potential ${\bf A}({\bf r})$ describes the effect of 
magnetic field. 

We assume that the graphene sheet is in the periodically alternating
magnetic field in $z$ direction. In our analytical calculations we
consider one-harmonic approach, $B({\bf r})=(B_0/3)\, {\rm
  Re}\sum_{i=1}^3 e^{i{\bf g}_i\cdot {\bf r}}$, where the in-plane
vectors ${\bf g}_i$ can be viewed as basic vectors of the triangular
lattice. In the numerical calculations we include more harmonics to
make the consideration realistic, preserving the triangular symmetry
of the field-imposed lattice.  One can use the gauge in which the
correspondent vector potential is also periodic, ${\bf A}({\bf
  r})={\rm Im}\sum _i{\bf a}_i\, e^{i{\bf g}_i\cdot {\bf r}}$, where
${\bf a}_i=(B_0/3g_i^2)\, ({\bf n}_0\times {\bf g}_i)$ and ${\bf n}_0$
is the unit vector along axis $z$ perpendicular to the plane.

We also assume that the period $a$ of alternating field is much much
larger than the lattice constant of graphene $a_0$. The large scale
field-induced triangular lattice determines the corresponding small
Brillouin zone in the inverse space (Fig.~1b), and we denote the
symmetry points of this zone by $K$, $M$, etc.  It should be stressed
that these points are not related in any way to the symmetry points of
the Brillouin zone of graphene, which we denote like $\mathcal{K}$
and $\mathcal{K}'$ (Fig.~1).

\section{Perturbation theory}

Let us assume $\Delta $ very small and positive, $\Delta >0$.  In the
case of weak periodic field, the term with ${\bf A}({\bf r})$ in (1)
is a weak perturbation.  We consider first the states related to the
Dirac point $\mathcal{K}$.  For ${\bf A}=0$, the unperturbed solution
for the low-energy state in the $\Gamma $ point with $\varepsilon >0$
is
\begin{eqnarray}
\label{2}
\psi ^{(0)}_{\Gamma ,+1}({\bf r})
=\frac1{\sqrt{\mathcal{S}}}\,
\left( \begin{array}{c} 1 \\ 0 \end{array}\right) ,
\hskip0.2cm
\psi ^{(0)}_{\Gamma ,-1}({\bf r})
=\frac1{\sqrt{\mathcal{S}}}\,
\left( \begin{array}{c} 0 \\ 1 \end{array}\right) ,
\end{eqnarray} 
where $\mathcal{S}$ the sample surface, and
index $\pm 1$ refers to the states in the first low-energy
energy band with positive and negative energy, respectively.  Assuming
$\Delta $ very small, for any other point $k\ne 0$ the solution is
\begin{eqnarray}
\label{3}
\psi ^{(0)}_{{\bf k},\pm 1}({\bf r})\simeq
\frac{e^{i{\bf k}\cdot {\bf r}}}{\sqrt{2\mathcal{S}}}
\left( \begin{array}{c}1 \\ \pm (k_x+ik_y)/k \end{array}\right) ,
\end{eqnarray}
corresponding to the states with energy $\varepsilon _{{\bf k},\pm
  1}\simeq \pm vk$.

The Hamiltonian of interaction is $\mathcal{H}_{int}=-\frac{ev}{c}\,
\bta \cdot {\bf A} $.  Matrix elements of the periodic field between
the states $\psi ^{(0)}_{{\bf k}\alpha }$ and $\psi ^{(0)}_{{\bf
    k'}\beta }$ are nonzero only for ${\bf k}-{\bf k'}={\bf g}$, where
${\bf g}$ is one of the vectors $\pm {\bf g}_1$, $\pm {\bf g}_2$, $\pm
{\bf g}_3$.  Correspondingly, the weak periodic field mostly affects
the states at the Brillouin zone edge, which have close in energy
counterparts, $\varepsilon ^{(0)}_{{\bf k}\alpha } \simeq \varepsilon
^{(0)}_{{\bf k}-{\bf g}_i,\alpha }$.  Then for the state at the
$\Gamma $ point $\psi _\Gamma ({\bf r})\simeq \psi _\Gamma ^{(0)}({\bf
  r})$, whereas the wave functions near the points $M$ and $K$ of
reciprocal lattice can be presented as a superposition of unperturbed
functions, $\psi _{\bf k}({\bf r})=c_0\psi ^{(0)}_{\bf k}({\bf r})
+\sum _ic_i\psi ^{(0)}_{{\bf k}-{\bf g}_i}({\bf r})$.

For the lower-in-energy states with $\varepsilon >0$ in $M$ and
$K$ points we find
\begin{eqnarray}
\label{4}
\psi _{M,+1}({\bf r})\simeq 
\frac{e^{i{\bf k}_M\cdot {\bf r}}}{2\sqrt{\mathcal{S}}}
\left[
\left( \begin{array}{c} 1 \\ 1 \end{array} \right)
+e^{-i{\bf g}_3\cdot {\bf r}}
\left( \begin{array}{c} 1 \\ -1 \end{array} \right) \right] .
\\
\psi _{K,+1}({\bf r})\simeq 
\frac{e^{i{\bf k}_K\cdot {\bf r}}}{\sqrt{6\mathcal{S}}}
\left[
\left( \begin{array}{c}1 \\ \frac{\sqrt{3}+i}2 \end{array}\right) 
+e^{-i{\bf g}_2\cdot {\bf r}}
\left( \begin{array}{c}1 \\ -i \end{array}\right) 
\right. \nonumber \\ \left.
+e^{-i{\bf g}_3\cdot {\bf r}}
\left( \begin{array}{c}1 \\ \frac{-\sqrt{3}+i}2 \end{array}\right) 
\right] .
\end{eqnarray}
Using these functions we calculate the Berry phase $\gamma _C$ along the contour
$\Gamma $-$M$-$K$-$\Gamma $, shown in Fig.~1b, as
\begin{eqnarray}
\label{5}
\gamma _C
=\arg \left< \Gamma |M\right>\left< M|K\right> \left< K|\Gamma \right> =0,
\end{eqnarray}
Thus, for the contour along the whole Brillouin zone and
the first energy band we obtain
$\gamma _{+1}=12\gamma _C=0$. Correspondingly, the Chern number of the $(+1)$ band
is ${\rm Ch}_{+1}=\gamma _{+1}/2\pi =0$.
Using the same method for the negative energy band $(-1)$, we find 
$\gamma _{-1}=-2\pi $ and ${\rm Ch}_{-1}=-1$.

\begin{figure}[ht]
\includegraphics[scale=1.1,angle=0]{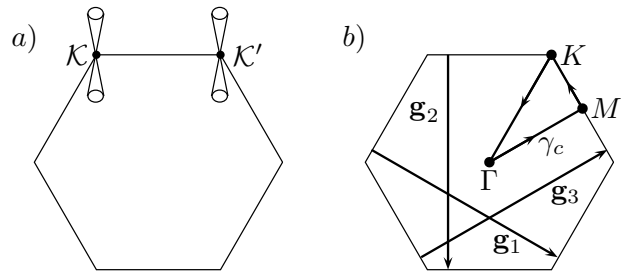}
\caption{(a) Brillouin zone of graphene with two nonequivalent Dirac points;
(b) Brillouin zone corresponding to a single Dirac point of graphene
in a large-scale periodic field of triangular symmetry. 
Contour $\gamma _c$ in the $k$-space.}
\label{fig:1}
\end{figure}

We should also take into account the contribution of the another
non-equivalent Dirac point $\mathcal{K}'$. For this point the
Hamiltonian (1) differs by the opposite sign before $\partial _x$
and $A_x$. Calculating the wavefunctions in this case and using (6) we
can find for the positive energy band ${\rm Ch}_{+1}=-1$ and for
negative band ${\rm Ch}_{-1}=0$.

If we take $\Delta <0$, we obtain for the $\mathcal{K}$ point ${\rm
  Ch}_{+1}=-1$ and ${\rm Ch}_{-1}=0$. Correspondingly, for the
$\mathcal{K}'$ point the results are ${\rm Ch}_{+1}=0$ and ${\rm
  Ch}_{-1}=-1$.

\section{Symmetry arguments}

These results can be understood using the symmetry of Hamiltonian (1).
Indeed, the transition from $\mathcal{K}$ to $\mathcal{K}'$ with
simultaneous reversion of the sign of $\Delta $ is related to the
symmetry under the unitary transformation (rotation in sublattice
space) 
\begin{eqnarray}
\label{7}
\tau _y^{-1}\mathcal{H}^{\mathcal{K}}(\Delta )\tau _y
=\mathcal{H}^{\mathcal{K}'}(-\Delta ).
\end{eqnarray}
Such transformation applied
to the wavefunctions does not change the Berry phase (6).  Therefore,
the corresponding Chern numbers should be equal, 
\begin{eqnarray}
\label{8}
{\rm Ch}_n(\mathcal{K},\Delta )={\rm Ch}_n(\mathcal{K}',-\Delta ).
\end{eqnarray}
Besides, changing the sign of field (i.e., ${\bf A}\to -{\bf A}$) in
(1) with simultaneous spatial inversion and reversion of the sign of
$\Delta $ is equivalent to the change of sign of Hamiltonian
\begin{eqnarray}
\label{9}
\mathcal{H}({\bf r},{\bf A},\Delta )
=-\mathcal{H}(-{\bf r},-{\bf A},-\Delta ),
\end{eqnarray}
which corresponds to the electron-hole symmetry.
It can be also understood as a symmetry to the
pseudotime inversion, $\tilde{T}=i\tau _y\mathcal{R}$, where
$\mathcal{R}$ is the complex conjugation operator
\begin{eqnarray}
\label{10}
\tilde{T}^{-1}\mathcal{H}({\bf A},\Delta )\tilde{T}
=\mathcal{H}(-{\bf A},-\Delta ).
\end{eqnarray}
The complex conjugation operator acting on the
wavefunctions changes the sign of the Berry phase, and we obtain 
\begin{eqnarray}
\label{11}
{\rm Ch}_n({\bf A},\Delta )=-{\rm Ch}_n(-{\bf A},-\Delta ).
\end{eqnarray}
Also, there is a symmetry with respect to $\tau _z$ transformation:
\begin{eqnarray}
\label{12}
\tau _z^{-1}\mathcal{H}(\Delta )\, \tau _z
=-\mathcal{H}(-\Delta )
\end{eqnarray}
leading to ${\rm Ch}_n(\Delta )={\rm Ch}_{-n}(-\Delta )$.

Using these properties, we can analyze what happens when the parameter
$\Delta $ changes sign from negative to positive. It will helps us to
determine properly the AHE for the case of $\Delta =0$.

\section{Anomalous Hall effect}

We start from relation between the Chern numbers and AHE.  For the
contribution of points $\mathcal{K,K'}$ to the
off-diagonal conductivity, in the case when the chemical potential
$\mu $ is in the gap, one can use the standard formula
\cite{nagaosa10}
\begin{equation}
\label{13}
\sigma_{xy}=\frac{e^2}{\hbar}\sum_{{\bf k}n}
f(\varepsilon_{{\bf k}n})
\Omega_{{\bf k}n},
\end{equation}
where $f(\varepsilon )$ is the Fermi-Dirac distribution function,
$\Omega_{{\bf k}n}=\nabla_{\bf k}\times \bmatA
_{{\bf k}n}$ is the Berry curvature, and 
$\bmatA _{{\bf k}n}=-i\langle {\bf k}n|\nabla_{\bf k}|{\bf k}n\rangle$ 
is the gauge connection. The energy spectrum is shown in Fig.~2 for
different values of magnetic field (the discussion is in Sec.~VII).  

The sum over $n$ in Eq.~(13) formally includes infinite number of energy bands
with negative energies of electrons. However, one can show
that in the model of Eq.~(1) the contribution of all the bands with 
$\varepsilon _{{\bf k}n}<0$ to the AHE is equal to zero. The reason is
that if we take $\Delta \ne 0$ and the chemical potential $\mu =0$ than the sum 
of Chern numbers for the $\mathcal{K}$ and $\mathcal{K}'$ Dirac points
does not change with changing periodic field from $B({\bf r})$ to $-B({\bf r})$. 
Indeed, any gradual variation of the field amplitude from $B_0$ to $-B_0$ does not change
the Chern numbers because, as one can see from the band structure calculations,
this does not produce any field-induced band crossings. This does not depend
on the value and sign of $\Delta $ and therefore is also valid in the limit 
of $\Delta \to 0$. 
Since the sum of Chern numbers and the resulting AHE at $\mu =0$
is invariant with respect to field inversion, then, due to the antisymmetry
of AHE to the field, it should be zero. 
Thus, in the vacuum state with $\mu =0$ there is no Hall effect in graphene.
    
If the chemical potential $\mu \ne 0$ and is located somewhere within the gap, 
the AHE is not necessary
zero and can be calculated from the following formula, which gives
the quantized values of the Hall conductivity
\begin{eqnarray}
\label{14}
\sigma_{xy}=\frac{e^2}{h}\left( {\sum_{n}}^\prime \text{Ch}_n
-\sum _{n<0}\text{Ch}_n\right) .
\end{eqnarray}
Here ${\rm Ch}_n=\frac1{2\pi }\int d^2{\bf k}\; \Omega_{{\bf k}n}$
is the Chern number of the $n$-th energy band, and the first sum
in (14) runs over the occupied energy bands. 

Using the obtained results for the Chern numbers we find that the
contribution from the $\mathcal{K}$ point changes (in units of
$e^2/h$) from $-1$ to 0 when the parameter $\Delta $ changes from
negative to positive values.  Similarly, the contribution from the
$\mathcal{K}'$ switches from 0 to $-1$ when $\Delta $ changes from
$\Delta <0$ to $\Delta >0$.  The sum of contributions from both
$\mathcal{K}$ and $\mathcal{K}'$ is equal to $-1$ for any sign of
$\Delta $, and the sign of $\Delta $ determines, which of the points
$\mathcal{K}$ or $\mathcal{K}'$ is contributing to the AHE.  But in
the point of $\Delta =0$, which corresponds to the gapless model of
graphene, the contributions from $\mathcal{K}$ and $\mathcal{K}'$ are
{\it exactly the same}.  Therefore, we come to conclusion that at
$\Delta =0$ each of these points gives the {\it fractional} value of
quantized Hall conductivity equal to $-1/2$.

\section{Gapless Dirac model}

Let us consider now the Hamiltonian (1) with $\Delta =0$. It has the
symmetry $\tau _z^{-1}\mathcal {H}\tau _z=-\mathcal{H}$, which means
that if $\psi _n$ is the eigenfunction of $\mathcal{H}$ with energy
$\varepsilon _n$ then $\tau _z\psi _n$ corresponds to the state with
energy $-\varepsilon _n$.  The $\tau _z$ symmetry also means that the
Berry phase calculated for any positive $(+n)$-band is exactly the
same as for the $(-n)$ band.
   
\begin{figure}[ht]
\centering
\includegraphics[scale=1,angle=0]{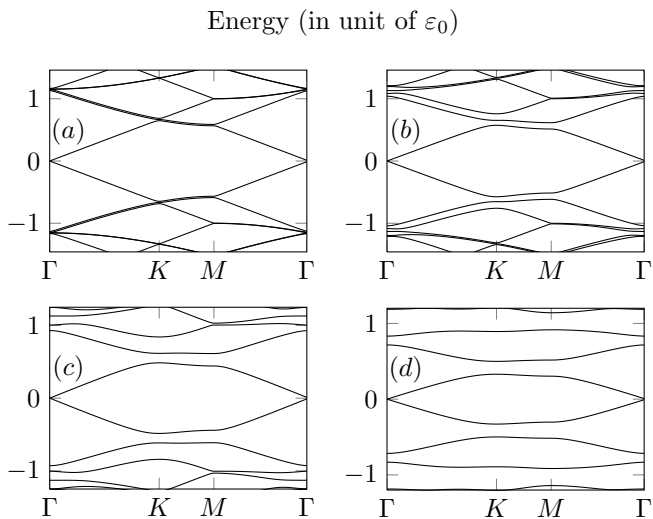}
\caption{The electron band structure of graphene ($\Delta =0$)
for different intensities of the field:
(a) $\alpha =0.05$, (b) $\alpha =0.2$, (c) $\alpha =0.5$, (d) $\alpha =1.0$.}
\label{fig:2}
\end{figure}   
   
For any ${\bf k}$-point including $\Gamma $ we use the unperturbed
wave functions (3). One can calculate the Berry phase $\gamma _C$
along the contour of Fig.~1b, excluding the $\Gamma $ point and using
\begin{eqnarray}
\label{15}
\gamma _C=\arg \left< \Gamma |M\right>
\left< M|K\right> \left< K|\Gamma '\right> \left< \Gamma '|\Gamma \right> ,
\end{eqnarray}
where $\Gamma $ and $\Gamma '$ correspond to the states with ${\bf
  k}\simeq 0$ along the $\Gamma $-$M$ and $\Gamma $-$K$ lines,
respectively.  It gives us for the $\mathcal{K}$ point and
$\varepsilon >0$, $\gamma _C=-\pi /12$.  The contour around the
Brillouin zone gives us $\gamma _{+1}=12\gamma _C+\gamma _0$, where
$\gamma _0$ is the Berry phase for a small contour around the $\Gamma
$ point.  The calculation gives $\gamma _0=\pi $. As a result we 
find the integer Chern number ${\rm Ch}_{+1}=0$. However, this Chern number
does not determine the Hall conductivity corresponding to Eq.~(14)
because the contribution of $\gamma _0$ should be excluded from
consideration as it does not depend on the external field.  Thus, to
calculate the AHE we should use the Berry phase without enclosing
$\Gamma $ point, and we finally find that the contribution to AHE from
any of the $\mathcal{K}$ or $\mathcal{K}$ points is equal to $-1/2$ in
accordance with what we found before.
     
\begin{figure}[ht]
\includegraphics[scale=1,angle=0]{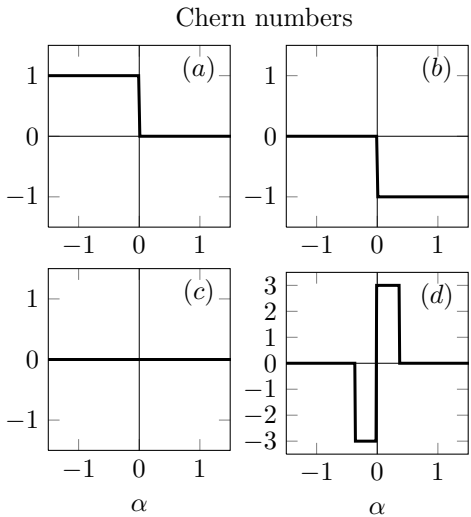}
\caption{Numerically calculated Chern numbers of the first positive energy 
bands for small $\Delta \ne 0$ as a function of $\alpha $:
(a) ${\rm Ch}_{+1}$ for $\Delta >0$, 
(b) ${\rm Ch}_{+1}$ for $\Delta <0$, 
(c) ${\rm Ch}_{+2}$,
(d) ${\rm Ch}_{+3}$.}
\label{fig:3}
\end{figure}

\begin{figure}[ht]
\includegraphics[width=7.5cm,angle=0]{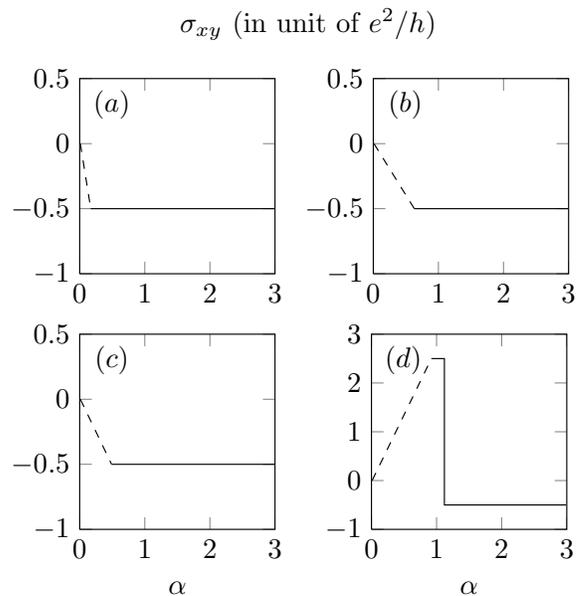}
\caption{Hall conductivity per one spin and one valley as a function of magnetic 
field for different values of electron density: (a) $n_0=4$, (b) $n_0=8$, 
(c) $n_0=12$, (d) $n_0=16$, where $n_0$ is the total number of electrons per 
elementary cell of graphene in periodic field.}
\label{fig:4}
\end{figure}

\section{Numerical calculations}

Using Hamiltonian (1) we performed numerical calculations of the
energy band structure for different intensity of the periodic
field. In these calculations we used the real numerically calculated
distribution of magnetic field created by the lattice of magnetic
nanorods, not restricting ourselves to harmonic
approximation.  The results are presented in Fig.~2 for different
values of the dimensionless parameter of intensity of periodic field
$\alpha $, defined as $\alpha=B_0 a^2 /4\pi \phi_0$, where $\phi_0$ is
the flux quantum. Figure 2 demonstrates that the energy gap between
the $(+1)$ and $(+2)$ bands appears at the field corresponding to
$\alpha >0.2$.

We also calculated numerically the Chern numbers as a function of the
field intensity.  They are presented in Fig.~3 for the first three
bands.  It turns out that the results of perturbation theory for the
low-energy bands $n=\pm 1$ remains valid with increasing $\alpha $.
Figure 4 presents the Hall conductivity per one spin and one valley as
a function of concentration of electrons. The dashed line corresponds
to the unfilled energy band, where an additional contribution from the
Fermi surface should be taken into account \cite{nagaosa10} but we do
not consider this case.

\section{Estimation of parameters}

The results of numerical calculations show that the first gap between
the $(+1)$ and $(+2)$ bands appears at $\alpha >0.2$.  In the case of
iron nanorod lattice with a lattice parameter $a\simeq 150$~nm, we
find that the magnitude of $B_0$ at a distance about 15 nm from the
nanorod lattice is $B_0\simeq 0.2$~T \cite{taillefumier08} and that
$\alpha \simeq 0.2$, which is in good agreement with the criteria of
gap formation.

One can also estimate the electron density $n$ corresponding to the
insulating state when the chemical potential $\mu$ is located in the
first gap. This state is characterized by one electron per elementary
cell, $n=4/\mathcal{S}_0$, where $\mathcal{S}_0=a^2\sqrt{3}/2$ and
factor 4 takes into account degeneracy with respect to spin and valley
in graphene. Using the relation $n=\mu ^2/\pi v^2$, where
$v/\hbar \simeq 1.5\times 10^6$~m/s is the electron velocity, we find $\mu
\simeq 40$~meV. All necessary parameters are quite reachable for the
technology of graphene.

One can estimate the characteristic energy $\varepsilon_0=h v/a\simeq
40$~meV for $a=150$~nm.  From the band structure calculations we
expect the gap to be of the order of $\varepsilon_0/6$. So, the gap is
$\simeq 90$~K. For the Zeeman splitting in graphene we have
$g\mu_BB_0\simeq 0.5$~K, so we expect that Zeeman splitting is
negligibly small for this choice of parameters.

\section{Discussion}

We found that the AHE related to a single Dirac point in
graphene is fractionally quantized, corresponding to quantum number
$-1/2$. The quantization of AHE takes place if the chemical potential
is located in the gap between the bands (+1) and (+2), which can be
easily achieved with the existing experimental technique.  The
fractional quantization is related to the Dirac points, which is the
monopole (for the $\mathcal{K}$ point) or antimonopole (for
$\mathcal{K}'$) in the ${\bf k}$ space, but the Berry phase related to
these points does not contribute to the AHE.
Correspondingly, the measured AHE in graphene with two nonequivalent
Dirac points and two spin orientations is integer for the same
parameters, $\sigma _H=ne^2/\hbar $ with $n=-2$.  

Our consideration of the AHE is essentially based on the statement that in 
the clean case with $\mu =0$ the AHE is zero. 
In the above-studied model with $\Delta \ne 0$, it is related
to the invariance of Hamiltonian and Chern numbers of the filled bands
with respect to the inversion of field, which contradicts to antisymmetry of AHE. 
It does not depend on sign of $\Delta $ and
leads to the zero vacuum AHE in the limiting case of $\Delta =0$ in 
graphene.   

In this connection, it is important to compare our results with the
result of Haldane's model \cite{haldane88}, in which the AHE can be
nonzero even in the vacuum state. In the Haldane's model the periodic field 
has the same periodicity as the honeycomb lattice, and the model includes additional
next-neighbor hoppings. Then the next-neighbor hopping integral in such a fast alternating 
field acquires nonzero phase factor, which leads to the brake of
electron-hole symmetry. Obviously, this model is not invariant with respect
to the field inversion, $B({\bf r})\to -B({\bf r})$.
As a result, our arguments about vanishing of AHE in the ground state of graphene 
are not applicable to the model of Ref.~[\onlinecite{haldane88}].   

\begin{figure}[t]
\vspace*{-1cm}
\includegraphics[scale=0.38,angle=0]{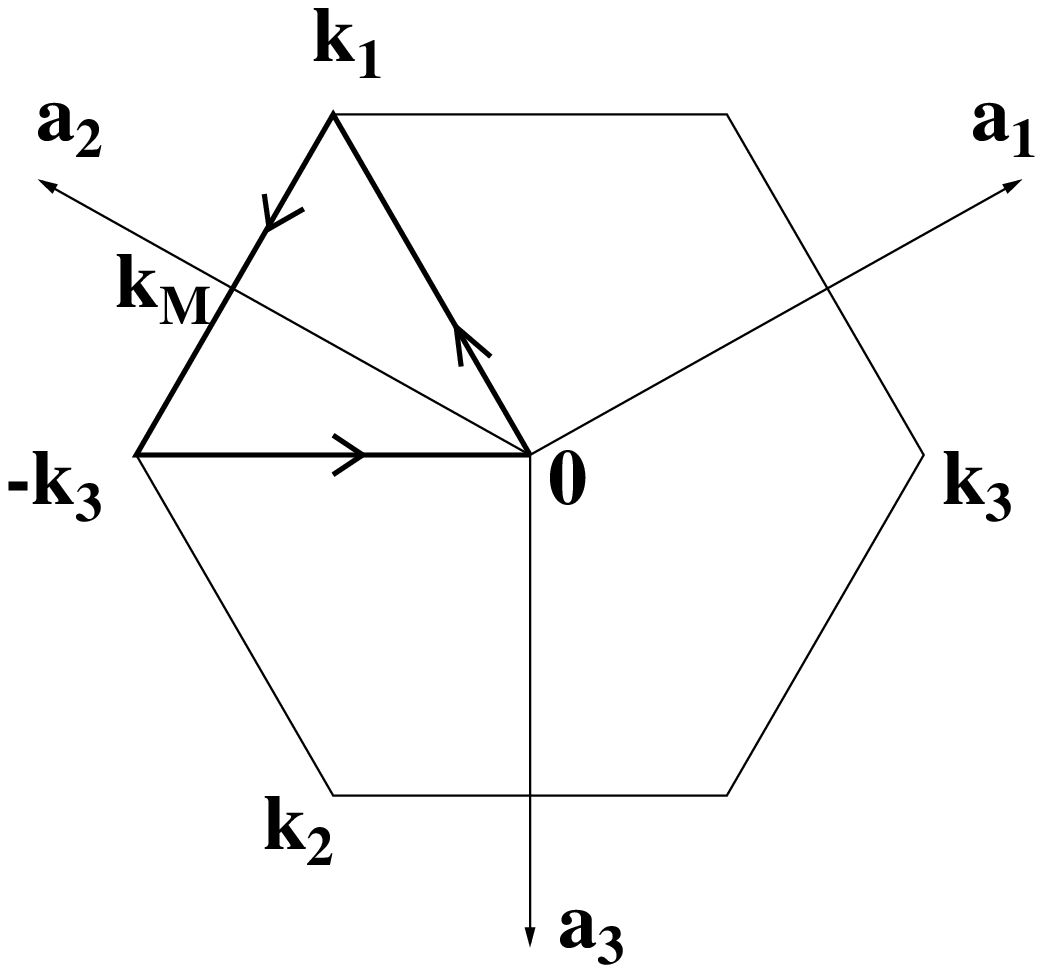}
\hspace*{-1cm}
\includegraphics[scale=0.38,angle=0]{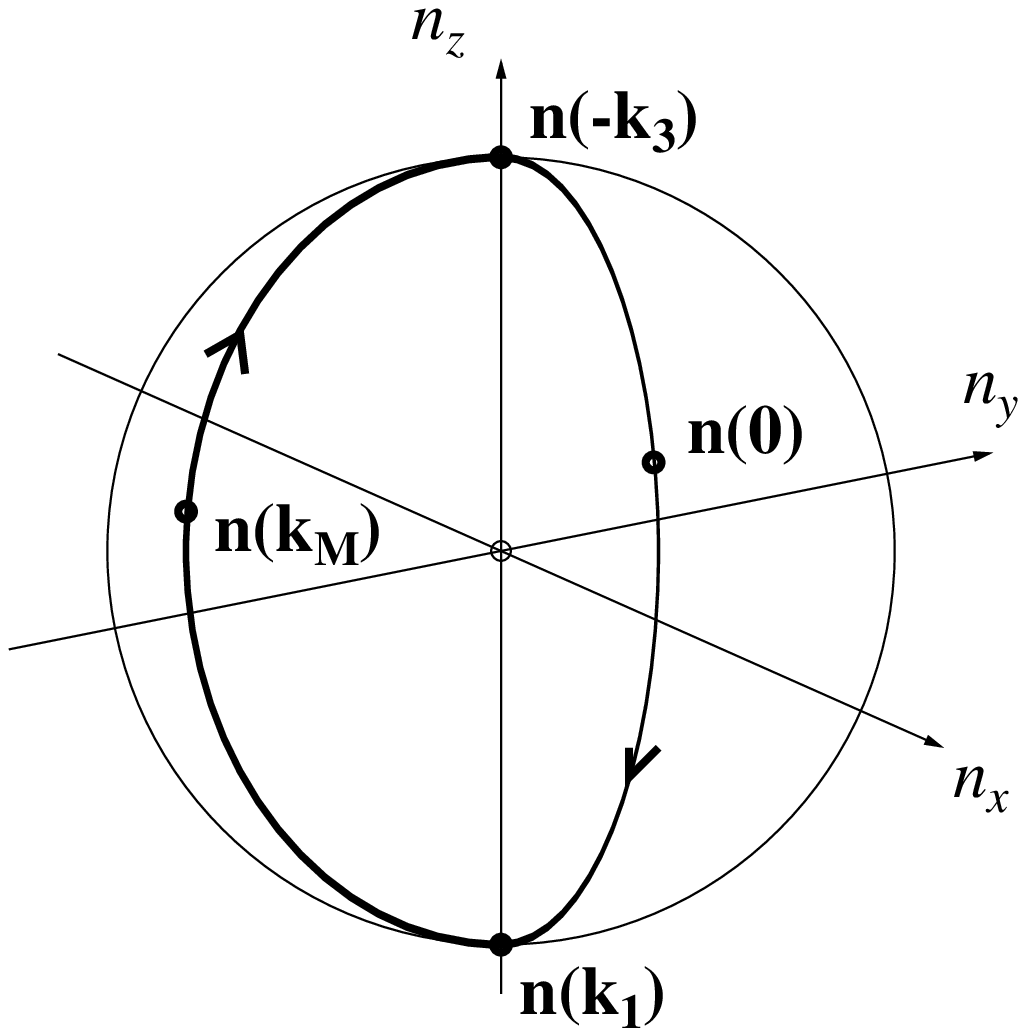}
\caption{Brillouin zone and Berry sphere of the Haldane's model in the case of
$\Delta <3\sqrt{3}t_2\sin \phi $.}
\label{fig:4}
\end{figure}

To clarify this point we reconsider the Haldane's model by calculating the Hall
conductivity as the Berry phase of electrons in the lower subband along the path 
encompassing the Brillouin zone of the honeycomb lattice. 
The Hamiltonian of Ref.~[\onlinecite{haldane88}] can be presented as
\begin{eqnarray}
\label{16}
\mathcal{H}_H=\varepsilon ({\bf k})+\lambda \bsig \cdot {\bf n}({\bf k})
\end{eqnarray}
where the unit vector ${\bf n}({\bf k})$ parametrizes $2\times 2$ pseudospin 
Hamiltonian. Here we denoted 
\begin{eqnarray}
\label{17}
&&\varepsilon ({\bf k})=2t_2\cos \phi \sum _i\cos ({\bf k}\cdot {\bf b}_i),
\\
&&n_x({\bf k})=(t_1/\lambda )\sum _i\cos ({\bf k}\cdot {\bf a}_i) ,
\\
&&n_y({\bf k})=(t_1/\lambda )\sum _i\sin ({\bf k}\cdot {\bf a}_i) ,
\\
&&n_z({\bf k})=[\Delta -t_2\sin \phi \sum _i\sin ({\bf k}\cdot {\bf b}_i)]/\lambda ,
\end{eqnarray}
where the lattice vectors ${\bf a}_i$ and ${\bf b}_i$ are defined in [\onlinecite{haldane88}] 
and the parameter $\lambda $ is introduced to provide $|{\bf n}|$=1.  
The parameter $t_2$ is the next-neighbor hopping integral and the phase $\phi $
is due to this hopping in the field with the periodicity of crystal lattice. 

The Berry phase for the motion along the path 
$(0,\, {\bf k}_1,\, {\bf k}_M,\, -{\bf k}_3,\, 0)$
is equal to the spherical angle of the mapping ${\bf k}\to {\bf n}({\bf k})$
to the Berry sphere (see Fig.~5).
Using (17)-(20) we find that in the case of $\Delta >3\sqrt{3}t_2\sin \phi $ the
vector ${\bf n}(-{\bf k}_3)$ points down coinciding with ${\bf n}({\bf k}_1)$,
which leads to the Berry phase $\gamma _C=0$ and Chern number ${\rm Ch}=0$. 
In the opposite case of
$\Delta <3\sqrt{3}t_2\sin \phi $, vector ${\bf n}(-{\bf k}_3)$ points up
(as shown in Fig.~5) leading to $\gamma _C=-\pi /3$ and, correspondingly, 
we obtain in this case ${\rm Ch}=-1$.
The same result follows from the explicit calculation of the Berry
phase using relation $\gamma _C={\rm arg}\, \left< 0|{\bf k}_1\right>
\left< {\bf k}_1|{\bf k}_M\right> \left< {\bf k}_M|-{\bf k}_3\right>
\left< -{\bf k}_3|0\right> $. This is the result of Ref.~[\onlinecite{haldane88}] 
found by geometrical method.

When the constant of field lattice $a$ is much larger than the lattice
constant of graphene $a_0$, the electron-hole and inversion symmery of graphene 
should be 
restored. In this limit, the field flux through an elementary cell
is nonzero breaking locally the inversion symmetry but the number
of cells with positive and negative flux is equal at the scale $\l \gg a$.      
Besides, in any realistic structure, the spatial fluctuations 
of field periodicity are much larger than the lattice constant, which
makes it impossible to provide the compatibility of two different lattices.  

We believe that the
quantized Hall effect in the structure with graphene on top of the
magnetic nanolattice can have important practical applications similar
to the usual quantum Hall effect in magnetic field.

\section*{Acknowledgments}

This work is supported by the FCT Grant PTDC/FIS/70843/2006 in
Portugal and by the National Science Center in Poland in years 2011 -- 2014.


\begin{thebibliography}{99}

\bibitem{novoselov04}
K. S. Novoselov, A. K. Geim, S. V. Morozov, D. Jiang, Y. Zhang, S. V. Dubonos, 
I. V. Grigorieva, and A. A. Firsov, Science {\bf 306}, 666 (2004).

\bibitem{novoselov05}
K. S. Novoselov, 
A. K. Geim, S. V. Morozov, D. Jiang, M. I. Katsnelson,
I. V. Grigorieva, S. V. Dubonos, and A. A. Firsov, 
Nature (London) {\bf 438}, 197 (2005).

\bibitem{geim07}
A. K. Geim and K. S. Novoselov, Nature Mater. {\bf 6}, 183 (2007).

\bibitem{neto09}
A. H. Castro Neto, 
F. Guinea, N. M. R. Peres, K. S. Novoselov, and A. K. Geim, 
Rev. Mod. Phys. {\bf 81}, 109 (2009).

\bibitem{peres10}
N. M. R. Peres, J. Phys: Cond. Matter. {\bf 21}, 323201 (2010).

\bibitem{zhang05}
Y. Zhang, Y. W. Tan, H. L. Stormer, and P. Kim, Nature {\bf 438}, 201 (2005).

\bibitem{novoselov07}
K. S. Novoselov, Z. Zhang, Y. Zhang, S. V. Morozov, H. L. Stormer,
U. Zeitler, J. C. Maan, G. S. Boebinger, P. Kim, and A. K. Geim,
Science {\bf 315}, 1379 (2007).

\bibitem{taillefumier08}
M. Taillefumier, V. K. Dugaev, B. Canals, C. Lacroix, and P. Bruno,
\prb {\bf 78}, 155330 (2008).

\bibitem{bruno04}
P. Bruno, V. K. Dugaev, and M. Taillefumier, \prl {\bf 93}, 096806 (2004).

\bibitem{nielsch01}
K. Nielsch, 
R. B. Wehrspohn, J. Barthel, J. Kirschner, U. G\"osele, S. F. Fischer, and H.~Kronm\"uller, 
Appl. Phys. Lett. {\bf 79}, 1360 (2001). 

\bibitem{rossler06}
U. K. R\"o\ss ler, A. N. Bogdanov, and C. Pfleiderer, Nature {\bf 442}, 797 (2006).

\bibitem{yu10}
X. Z. Yu, 
Y. Onoze, N. Kanazawa, J. H. Park, J. H. Han, Y. Matsui, N. Nagaosa, and Y. Tokura, 
Nature {\bf 465}, 901 (2010).

\bibitem{nagaosa10}
N. Nagaosa,
J. Sinova, S. Onoda, A. H. MacDonald, and N. P. Ong,
\rmp {\bf 82}, 1539 (2010). 

\bibitem{snyman09}
I. Snyman, \prb {\bf 80}, 054303 (2009).


\bibitem{haldane88}
F. D. M. Haldane, \prl {\bf 61}, 2015 (1988).

\end{thebibliography}
\end{document}